\documentclass[fleqn,usenatbib]{mnras}

\usepackage{newtxtext,newtxmath}

\usepackage[T1]{fontenc}
\usepackage{xcolor}
\DeclareRobustCommand{\VAN}[3]{#2}
\let\VANthebibliography\thebibliography
\def\thebibliography{\DeclareRobustCommand{\VAN}[3]{##3}\VANthebibliography}

\usepackage{graphicx}	
\usepackage{amsmath}	
\usepackage{comment}

\title[Virialized EoS]{Virialized equation of state for warm and dense  stellar plasmas in proto-neutron stars and Supernova matter}

\author[D. Barba-González et al.]{
D. Barba-González\thanks {E-mail: david.barbag@usal.es}
C. Albertus\thanks{E-mail: albertus@usal.es}
and M.A. Pérez-García\thanks{E-mail: mperezga@usal.es}
\\
Department of Fundamental Physics and IUFFyM,\\ Universidad de Salamanca, Plaza de la Merced s/n E-37008, Salamanca (Spain)
}

\date{Accepted XXX. Received YYY; in original form ZZZ}

\pubyear{2024}

\begin{document}
\label{firstpage}
\pagerange{\pageref{firstpage}--\pageref{lastpage}}
\maketitle

\begin{abstract} 
We present microscopic Molecular Dynamics simulations including the efficient Ewald sum procedure to study warm and dense stellar plasmas consisting of finite-size ion charges immerse in a relativistic neutralizing electron gas.  For  densities typical of Supernova matter and crust in a proto-neutron star, we select a representative single ion composition and obtain the virialized equation of state (vEoS). We scrutinize the finite-size and screening corrections to the Coulomb potential appearing in the virial coefficients $B_2, B_3$ and $B_4$ as a function of temperature. In addition, we study the thermal heat capacity at constant volume, $C_V$, and the generalized Mayer's relation i.e. the difference $C_P-C_V$ with $C_P$  being the heat capacity at constant pressure, obtaining clear features signaling the onset of the liquid-gas phase transition. Our findings show that microscopic simulations reproduce the discontinuity in $C_V$, whose value lies between that of idealized gas and crystallized configurations. We study the pressure isotherms marking  the boundary of the metastable region before the gaseous transition takes place. The resulting vEoS displays a behaviour where effective virial coefficients include extra density dependence showing a generalized density-temperature form. As an application we parametrize pressure as a function of density and temperature under the form of an artificial  neural network showing the potential of machine learning for future regression analysis in more refined multicomponent approaches. This is of interest to size the importance of these corrections in the liquid-gas phase transition in warm and dense plasma phases contributing to the cooling behaviour of early Supernova phases and proto-neutron stars. 
\end{abstract}

\begin{keywords}
plasmas, (stars:) supernovae: general, stars: neutron, equation of state, dense matter
\end{keywords}



\section{Introduction}

\label{intro}
In nature there are several astrophysical scenarios where the relevant dynamics of matter can be described in terms of screened ion systems. Examples of them are low density warm Supernova matter or the outer layers of proto Neutron Stars (NSs). Ions are immerse in a neutralizing degenerate electron Fermi sea formed at densities beyond $10^6 \,\rm {g}/{cm^3}$, as a result of electrons being  stripped off atoms and screening the otherwise nude Coulomb interaction. While at sufficiently low temperatures, below the melting $T< T_{\rm melt}$, matter arranges in a bcc, fcc or hcp crystallized configuration, for densities below the neutron drip \citep{fcc_PhysRevE.103.043205}, at higher temperatures and/or densities matter adopts a fluid state undergoing for this a phase transition, see \cite{phasetransition_PhysRevE.49.5164}.
Such plasmas display, naturally, a complex composition \citep{2022MNRAS.513L..52C}, however typical approaches to their modelization have considered pure systems, i.e. a one component plasma (OCP) with a single ionic species for the sake of simplicity. However, in warm and dense stellar plasmas like the ones cited, a multicomponent (MCP) scenario is the most likely, see Fig. 1 in  \cite{snwam_10.1093/mnras/sty3468}. The MCP system arises due to chemical and electrostatic equilibrium at given thermodynamical conditions from an existing previous composition late in the Supernova aftermath. Thus in the more realistic picture, the composition displays some variability with the ion population showing some spread around a most frequent ion.
\begin{figure*}
    \centering
    \includegraphics[width=\textwidth]{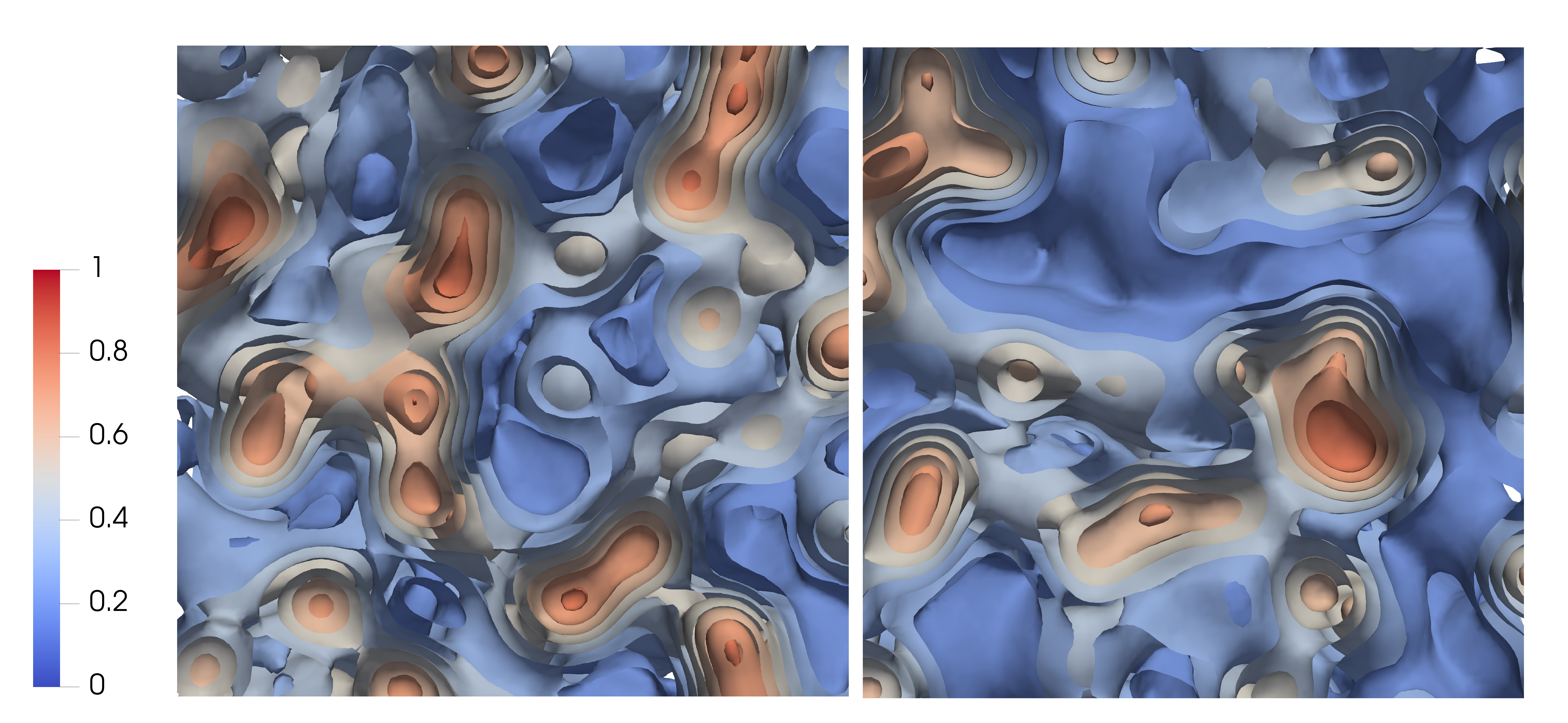}
    \caption{Density isosurfaces for the screened Coulomb ion OCP gas with $(Z,A)=(38,128)$ implemented with Ewald sums and finite ion charge spread at $k_BT = 8$ MeV as obtained with our MD code. We have rescaled from 0 to 1 the local densities considering (left panel) an average ionic density $n_{I}=1.1 \times 10^{-4}$ $ \rm fm^{-3}$ corresponding to $\Gamma_C= 20.08$ and  (right panel) $n_{I}=7 \times 10^{-6}$ $\rm fm^{-3}$ ($\Gamma_C=8.02$). In both cases the screening parameter is $\kappa=0.622$.}
    \label{fig:paraview}
\end{figure*}

Previous works at very low densities have provided  the finite-temperature EoS in a model independent way under the form of the virial EoS. In \cite{HOROWITZ200655} second virial coefficients were obtained for a system with neutrons, protons, and $\alpha$-particles along with experimental information on binding energies and phase shifts. \cite{oconnor_PhysRevC.75.055803} and \cite{mallik_PhysRevC.77.032201} added heavier species, obtaining as a result that they contribute substantially to the virial series for pressure. For light nuclei the quantum statistical model of \cite{ROPKE201370} extended the quasiparticle virial expansion to include arbitrary point-like clusters. An application of this to H plasmas has been recently developed \citep{ropke2023thermodynamic}.

The link between the macroscopic expression given by the virialized equation of state (vEoS) and the microscopic world was made in the 1930s by Uhlenbeck and Beth when they showed that the virial coefficients could be given in terms of integrals involving the interaction potential energy \citep{UHLENBECK1936729}.
Besides, in realistic cooling MCP scenarios in the partially accreted inner crust at $k_B T\lesssim1$ MeV ($k_{\mathrm{B}}$ is the Boltzmann constant) heat capacity at constant volume is a function of density, mostly detemined by ions as shown in \citep{potekhin}. Thus understanding how the crust forms from an originally warm dense plasma in a proto NS is interesting by itself. Note the densities involved are those below the neutron drip, $\rho_{ND}\sim 10^{11.6}$ $\rm g/cm^3$, being this value somewhat model dependent. The crust formation in a newly born NS proceeds from a  $k_B T\gtrsim 20$ MeV initial state in the aftermath of gravitational collapse to cool down to $k_B T\lesssim 0.1$ MeV in days to months. 

Here we focus on the properties of a plasma describing  the warm and dense Supernova matter in the regime of nuclear statistical equilibrium (NSE) where photo-disintegration and radiative captures shape the nuclear population or in early proto NS outer layers. 
Previous studies \citep{Ishizuka_2003} show the critical temperature may be very high $T_c\sim14$ MeV for lepton fraction $Y_L=(0.3-0.4)$, which is the ratio expected in actual supernova explosions, see \cite{oertelRevModPhys.89.015007}. These critical temperatures are much higher than those in neutrino-less Supernova matter and comparable to those in symmetric nuclear matter. The boiling points in those cases remain at $T_{boil}\sim 1$ MeV even at very low densities for all lepton fractions, see \cite{dinh} for a recent calculation.
From this the plasma composition is determined by Saha equations or extended NSE models involving nuclei, neutrons, protons, and $\alpha$ particles and when T cools down below some value, the composition frozens and is essentially unchanged, see a discussion in \cite{RADUTA2019252}. Generally speaking, to model the cooling behaviour in this complex system one should rely on screened ionic degrees of freedom and solve (ideally in radial coordinate, $r$) for local (non-redshifted) temperature $T(r)$, in a locally flat space-time.  Assuming no external mass transfer nor radiation the heat equation can be written as 
\begin{equation}
\rho C_{P} \frac{\partial T}{\partial t}-\nabla \cdot(\kappa_0 \nabla T)=\dot{Q}_{V}    
\label{heat}
\end{equation}
where $\rho$, $C_{P}$ and $\kappa_0$ are the mass density, heat capacity per unit volume at constant pressure and  thermal conductivity, respectively. $\dot{Q}_{V}$ is the volumetric emissivity accounting for heat transfer.

In this work we perform Molecular Dynamics (MD) simulations for a low density ion system whose composition is compatible with those obtained in the literature for NS crusts, see \cite{pearson2018} and \cite{murarka2022}, although we note there is some spread regarding the election of the single-ion species in the OCP description for same density range. In particular, we use the innermost ion of table 4 in \cite{pearson2018} and focus on ion densities in the interval  $n_I\in [2,7]\times 10^{-6}$ $\rm fm^{-3}$ within the OCP approximation with the ion species $(Z,A)=(38,128)$ having its charge spread under a finite-width gaussian distribution with zero spin. Accordingly, the temperature range that we will explore in our MD simulations is $k_BT\in \left[2,11\right]$ MeV, corresponding to a range of $\Gamma_C\in \left[3.84,32.06\right]$,  so that crystallization does not happen and liquid and gas phases appear. The screening is set at $\kappa=0.622$.  We will consider, relying on previous works \cite{Ishizuka_2003},  that dissolution of the nuclear clusterized entities in our neutral neutron rich plasma $Y_p\sim0.3$ does not take place either and, in this spirit, we extend our analysis to the relatively higher T range for the sake of completion. In that work the authors discussed that for systems with $Y_p\lesssim 0.35$ boiling temperatures $T_{boil}\lesssim 10$ MeV. In the same line as shown in \cite{dinh} clusterized degrees of freedom are still valid in the warm phases within the crust density range in the proto-NS, with compositions compatible to those in the inner crust of a catalyzed star.  

The simulations we present were performed using our original code $\rm USALMD_{Gauss-Ewald}$ using a multicore computing setup with efficient parallelization in \texttt{Fortran+OpenMP}, already presented in \cite{barba1}. In our runs all positions and velocities of ions are followed for at least $10^5 \,\rm {fm}/{c}$ to assure thermalization before extracting the thermodynamical properties of the system. The NVT ensemble is used, whereas we fix the volume by using periodic boundary conditions in an Ewald summation \citep{Ewald} paradigm calculation, and temperature is fixed by periodically rescaling the kinetic energy although its convergence to thermostats has been described in the past \citep{APGarcia2006}. 

The paper is organized in the following manner. In Section \eqref{section2} we present the  framework to study the screened realistic ion system at finite temperature. We obtain the pressure $P$ from the diagonal components of the general expression for the stress tensor arising from a Yukawa-like screened electrostatic interaction of finite-size ions with the Ewald construction. More in particular we obtain the pressure at finite temperature describing in detail the virial coefficients for the reference OCP and ion species $(Z,A)$ we use. From our simulated cases and to illustrate the potential of machine learning methods we perform a regression analysis and  present some useful expressions using artificial neural networks (ANN)  in the appendix \ref{appendix2} for the density-temperature phase space we sample. In Section \eqref{section3}, we further characterize the system describing the liquid-gas phase transition it undergoes by using the Ehrenfest criterion based on derivatives of the thermodynamical potential. We focus on constant pressure and volume heat capacities $C_P,C_V$ along with the generalized Mayer's relationship. We discuss how the non-ideal screened interaction ion system along with finite size  corrections distort the ideal gas result, that is safely recovered in the low density, high temperature regime. We discuss our findings comparing to some existing warm EoS in different frameworks and conclude in Section \eqref{section4}.   

\section{Realistic virial Equation of State for the warm and dense screened plasma}
\label{section2}

Finite temperature ion plasmas at a particular number density $n_I$ and temperature $T$ have been characterized by the dilute Coulomb fluid theory based on the parameter $\Gamma_C=\frac{Z^2}{lT}$ involving point-like particles. Here $Z$ is the charge of the ions and $l=\left(3/4\pi n_I\right)^{\frac{1}{3}}$ is the average interparticle distance between them.
However, at large electron mass densities $\rho_e>10^6$ $\rm g/cm^3$ typical in the NS outer crust it is expected that the ion plasma behaviour is better described by the Yukawa theory due to the electron screening of Coulomb interaction, as their degeneracy parameter i.e. Fermi energy to thermal energy ratio,  $\xi=E_{F,e}/k_BT \gg1$, signals their degenerate contribution.  The finite electron number density $n_e=Zn_I$  yields the ion system electrically charge neutral, also meaning that there is an additional parameter $\kappa=\lambda_e/l$,  with $\lambda_e$ the Thomas-Fermi screening length,  that shifts the crystallization point to lower temperatures in the Yukawa  treatment $\Gamma(\kappa)>\Gamma_C$ , see \citep{2002PhRvE..66a6404V,2015PhRvE..91b3108K}.  As it follows from the previous definition of $\kappa$ and for the OCP approximation we use, its density dependence drops off thus remaining fixed in the density range under study. On a more general  MCP system, however, this should not be the case. For $\kappa=0$ the Coulomb case is recovered. However, note that a more refined treatment using Jancovici screening instead of the usual Yukawa screening may distort this behaviour, for a discussion see \cite{potekhin}. Importantly, this description assumes ions as point-like particles with no spatial spread of the charge. When more realistic distributions are considered, an additional parameter $\eta=1/\sqrt{a}l$, where $a$ is a measure of the particle finite-width charge spread, it results a less energetically bound system, which can even be structurally affected shifting the crystallization point, see \cite{barba1}.

As we will discuss, the weaker interaction in Yukawa systems, and finite size effects to a less extent, impact the  EoS $P(\rho, T)$.
For a OCP description based on the most frequent ion at given density in the cold NS outer crust see \cite{fantina2020A&A...633A.149F} or \cite{dinh} for proto-neutron stars. Besides in the non-rigid crust approximation the low T behaviour of ions can impact the shear and bulk viscosities and mechanical properties with important consequences on thermal properties, bursts \citep{yak} and the ability of the star to sustain oscillation modes \citep{fantina18,suleiman22,gusakov,Samuelsson_2007}  just to cite some previous works.

As already presented in the context of realistic crust calculations in \cite{barba1} the Ewald sum procedure allows the very efficient calculation of high order correlations in systems with two-body interactions characterized by a non-zero  spatial scale $\lambda$. Examples of this have been already discussed in the literature, see \cite{salincaillol2003} or \cite{watanabe2003}.

In our calculation we consider the electron-screened Coulomb potential created at distance $|\vec{r}-\vec{r}_i|$ by the ith ion at position $\vec{r_i}$ i.e. the {\it Debye-Hückel potential}.  In the static and  point-like ion approximations the potential displays a Yukawa-like form
\begin{equation}
    \phi_i\left(\vec{r}\right)=\frac{Z_i}{|\vec{r}-\vec{r_i}|}e^{-\frac{|\vec{r}-\vec{r_i}|}{\lambda_e}},
    \label{phi1}
\end{equation}

\noindent where $\lambda_e\equiv \lambda$ is the Thomas-Fermi screening length  \begin{equation}
\lambda_e=\left(k^2_{F,e}+m_e^2\right)^{{-1}/{2}}\sqrt{\frac{\pi}{4\alpha k_{F,e}}} \sim \frac{1}{2 k_{F,e}}\sqrt{\frac{\pi}{\alpha}},
\end{equation}
and $k_{F,e}$ and $m_e$ are the electron Fermi momentum and mass, respectively. $\alpha$ is the fine structure constant. Electron number density is written as 
\begin{equation}
n_e=\sum_i Z_i n_{I,i}=\frac{k^3_{F,e}}{3 \pi^2}.
\end{equation}
for a set of ions $i=1,...N_I$, being in general a multicomponent system. 

In our scenario  thermal quantum effects are constrained by the fact that  they will not disturb the individual treatment of ions and collective of electrons as long as the thermal de Broglie wavelengths of the ions and electrons behave accordingly. Usually they are defined as $\lambda^{dB}_{\mathrm{I}}=\left(2 \pi \hbar^{2} / m_{\mathrm{I}} k_{B} T\right)^{1 / 2}$ and $\lambda^{dB}_{e}=\left(2 \pi \hbar^{2} / m_{e} k_{B} T\right)^{1 / 2}$, respectively. The quantum effects on ion motion are important either at $\lambda^{dB}_{\mathrm{I}} \gtrsim l_{\mathrm{I}}$ , where $l_I\equiv l$ or at $T \ll T_{\mathrm{p}}$, where $T_{\mathrm{p}} \equiv\left(\hbar \omega_{\mathrm{p}} / k_{B}\right) \lesssim 10^8$ K is the plasma temperature determined by the ion plasma frequency $\omega_{\mathrm{p}}=\left(4 \pi e^{2} n_{\mathrm{I}} Z^{2} / m_{\mathrm{I}}\right)^{1 / 2}$. Note that a fermion/boson ion idealized gas EoS will not be affected by such quantum corrections under this treatment, so that our MD treatment remains fully consistent for the $\rho-T$ ranges under study.

Following \cite{barba1} we will be considering a gaussian shape charge distribution for ions in the form  
\begin{equation}
 \rho_{Z_i,a_i} = Z_i\left(\frac{a_i}{\pi}\right)^{\frac{3}{2}}e^{-a_i r^2}   
\label{gauss1}
\end{equation}
with $a_i$ a characteristic width parameter depending on the ion $(Z,A)$. For details we refer to Appendix \ref{appendix1}. For the non point-like charges the ionic potential is obtained by solving the Poisson equation and is given by \cite{salincaillol2003}, and used by \cite{barba1} in the following form
\small
\begin{multline}
    \phi_{Zi,a_i}\left(\vec{r}\right)=\frac{Z_i}{2|\vec{r}-\vec{r_i}|} e^{\frac{1}{4 a_i \lambda_e^2}}\left[e^{-\frac{|\vec{r}-
\vec{r_i}|}{\lambda_e}}\mathrm{erfc}\left(\frac{1}{2\sqrt{a_i}\lambda_e} - \right.\right.\\ 
\left.\left. \sqrt{a_i}|\vec{r}-\vec{r_i}|\right)  -
e^{\frac{|\vec{r}-\vec{r_i}|}{\lambda_e}}\mathrm{erfc}\left(\frac{1}{2\sqrt{a_i}\lambda_e}+\sqrt{a_i}|\vec{r}-
\vec{r_i}|\right) \right],
\label{phi2}
\end{multline}

\normalsize
with erfc the complementary error function.

From the previous we can use the efficient Ewald procedure and separate the potential energy expression $U$ into short, long and self-interaction contributions. We detail the technical steps in Appendix \ref{appendix1}. Thus one can write $U=U_{\mathrm{short-range}}+  U_{\mathrm{long-range}}-  U_{\mathrm{self}}$ and similarly, the force over a particle $i$ be decomposed as a result of two-body contributions $\vec{F}_i=\sum_{j\neq i}^N \vec{F}_{ij,\mathrm{short-range}}+\vec{F}_{i,\mathrm{long-range}}$. These lengthy expressions include not only the real and screening corrections but also the finite-size ion gaussian charges within the Ewald summation frame. It is important to carefully implement the efficient sum, as the screening charge associated  inverse square width $\alpha_{\rm Ewald}$ also controls how the summation is distributed between the short-range and long-range parts. It has been fixed ensuring that the minimum image convention is sufficient in real space \citep{2013JChPh.139x4108H}.%

The energy density $\rho$ is obtained in the system as the contribution from the kinetic and potential parts 

\begin{equation}
\rho=\frac{1}{L^3} \left[ \sum_{i=1}^{N_I} \frac{1}{2}m_{I,i} \dot{\vec{r}}_{i \alpha} \dot{\vec{r}}_{i \beta} \delta_{\alpha \beta} +U \right],
\end{equation}

where $\delta_{\alpha \beta}$ is the Kronecker delta and $\alpha,\beta=1,2,3$ are tensorial indexes for the XYZ spatial coordinates. $N_I=n_I L^3$ is the number of ions considered in our system and $L^3 = V$ is the volume of the cubic simulation box. 

In order to obtain the pressure $P$ in the system from the Ewald construction  we first obtain the general stress tensor $\Pi_{\alpha \beta}^{\mathrm{tot}}$ tailored specifically to use the Ewald procedure under the screened Coulomb potential created by the gaussian ion sources.  Let us note that in the fluid phase the off-diagonal components are zero as there is no shear. 

From the stress tensor diagonal components we obtain the corresponding pressure  $P=\frac{1}{3} {\rm Tr} \,\Pi^{\rm tot}_{\alpha \beta}$. 
This second-rank symmetric tensor is obtained from contributions of the real and Fourier reciprocal $k$-space as follows 

\begin{equation}
\Pi_{\alpha \beta}^{\mathrm{tot}}=\frac{1}{L^3} \sum_{i=1}^{N_I} m_I \dot{\vec{r}}_{i \alpha} \dot{\vec{r}}_{i \beta}+\Pi_{\alpha \beta}.
\end{equation}  Subsequently, it can be put into the form  (see Appendix \ref{appendix1}) $\Pi_{\alpha \beta}=\Pi_{\alpha \beta}^{\mathrm{real}}+\Pi_{\alpha \beta}^{\mathrm{recip}}+\Pi_{\alpha\beta}^{\mathrm{volume}}$. As the interactions in real space may be written in a pairwise fashion, 
$\Pi_{\alpha \beta}^{\text {real }}$ can be evaluated directly from the real forces by employing the virial expression
\begin{equation}
\Pi_{\alpha \beta}^{\text {real }}=\frac{1}{2 V} \sum_{i,{j\ne i}=1}^{N_I}\left(F_{i j, \alpha}^{\text {real }} r_{i j, \beta}+F_{i j, \beta}^{\mathrm{real}} r_{i j, \alpha}\right).
\end{equation}

The reciprocal part is obtained  including the  widths $a_i$ and ${\alpha_{\rm Ewald}}$ associated to the gaussian charge spread and Ewald construction respectively. For point-like approaches see \citep{aguado,salincaillol2003} coincident with the stress tensor calculations in \cite{NoseKlein}

\begin{equation}
\begin{aligned}
\Pi^{\text {recip }}_{\alpha \beta}&=  \frac{1}{2}\sum_{\vec{k}\neq0}^{\infty}\sum_{i,j}^{N_I} \frac{4\pi Z_i Z_j}{V^2\left(k^{2}+\frac{1}{\lambda_e^2}\right)}e^{\frac{-k^{2}}{4}\left(\frac{1}{a_i}+\frac{1}{\alpha_{\rm Ewald}}\right)}e^{i\vec{k}\cdot\left(\vec{r}_i-\vec{r}_j\right)} \times \\
&\left\{\delta_{\alpha \beta}-2 \left[ \frac{1}{4}\left(\frac{1}{a_i}+\frac{1}{\alpha_{\rm Ewald}}\right)+\frac{1}{\left(k^2+\frac{1}{\lambda_e^2}\right)}\right]k_{\alpha} k_{\beta} \right\}.
\end{aligned}
\label{reciprocal}
\end{equation}

This expression is obtained by taking the derivatives of the total energy with respect to generic deformations in the simulation box.
The volume part of the tensor comes from the last term in $U_{\mathrm{short-range}}$ (Eq. \eqref{Ushort} in the Appendix). It does not depend on the particle positions but it depends on volume, so that it adds to the stress tensor in the same way the reciprocal part in Eq. \eqref{reciprocal} does, as the derivative of volume with respect to deformations is finite. Then 
\begin{equation}
    \Pi_{\alpha\beta}^{\mathrm{volume}}=-\frac{2\pi\delta_{\alpha\beta}}{V^2}\sum_{i=1}^{N_I}\sum_{j=1}^{N_I} Z_j \int_0^{\infty} r'^{2}\phi_{\rm{short-range}, \mathrm{i}} \left(r'\right)dr',
\end{equation}
where $\phi_{\rm{short-range}}$ is given by Eq. \eqref{phishort}.

We now proceed to express thermodynamic pressure in a suitable way in the screened finite-size ion system. In the point-like approach the {\it virial expansion} series adopts the generic form \citep{1902KNAB....4..125K}

\begin{equation}
\frac{P}{k_{\mathrm{B}} T}=\sum_{k=1}^{\infty} n_I^k B_k(T)
\label{virial}
\end{equation} 
with $B_1\equiv 1$ and $B_{n}$ with $n=2, 3, 4....,k$ being the T-dependent virial coefficients up to kth order. In particular, the $n=2$ term accounts for the contribution of  two-body interactions and has the form
\begin{equation}
B_{2}(T)=-\frac{1}{2 V} \iint f\left(r_{12}\right) \mathrm{~d} \mathbf{r}_{1} \mathrm{~d} \mathbf{r}_{2}
\label{b2}
\end{equation}
while effective three-body interactions are approximately described using the coefficient 

\begin{equation}
B_{3}(T)=-\frac{1}{3 V} \iiint f\left(r_{12}\right) f\left(r_{13}\right) f\left(r_{23}\right) \mathrm{d} \mathbf{r}_{1} \mathrm{~d} \mathbf{r}_{2} \mathrm{~d} \mathbf{r}_{3}.
\label{b3}
\end{equation}

\noindent and a similar expression for $B_{4}(T)$. Eqs. \eqref{b2}, \eqref{b3} are expressed in terms of Mayer cluster integrals \citep{mayer} involving  the quantity $f(r)=e^{-\frac{\phi_{Zi,a}(r)}{k_B T}}-1$ so that the larger $|f|$ is,  the more non-ideal the gas behaves. $B_n$ with $n\ge2$ give the corrections to the ideal behaviour.  Having obtained the $\sim \mathcal{O}(n^3_I)$ EoS it is possible to obtain the virial coefficient $B_4$, and so on. So, from the $B_2$, it is possible to compute approximately the full correction due to interactions in the procedure called Kirkwood superposition approximation. In our work we will restrict to fourth order vEoS. Additional refinements are indeed possible, so that coefficients in the expansion yield  information beyond a selected order in many-body correlations in the system, see \cite{oertelRevModPhys.89.015007} for a discussion.

\begin{figure}
\includegraphics[width=0.49\textwidth]{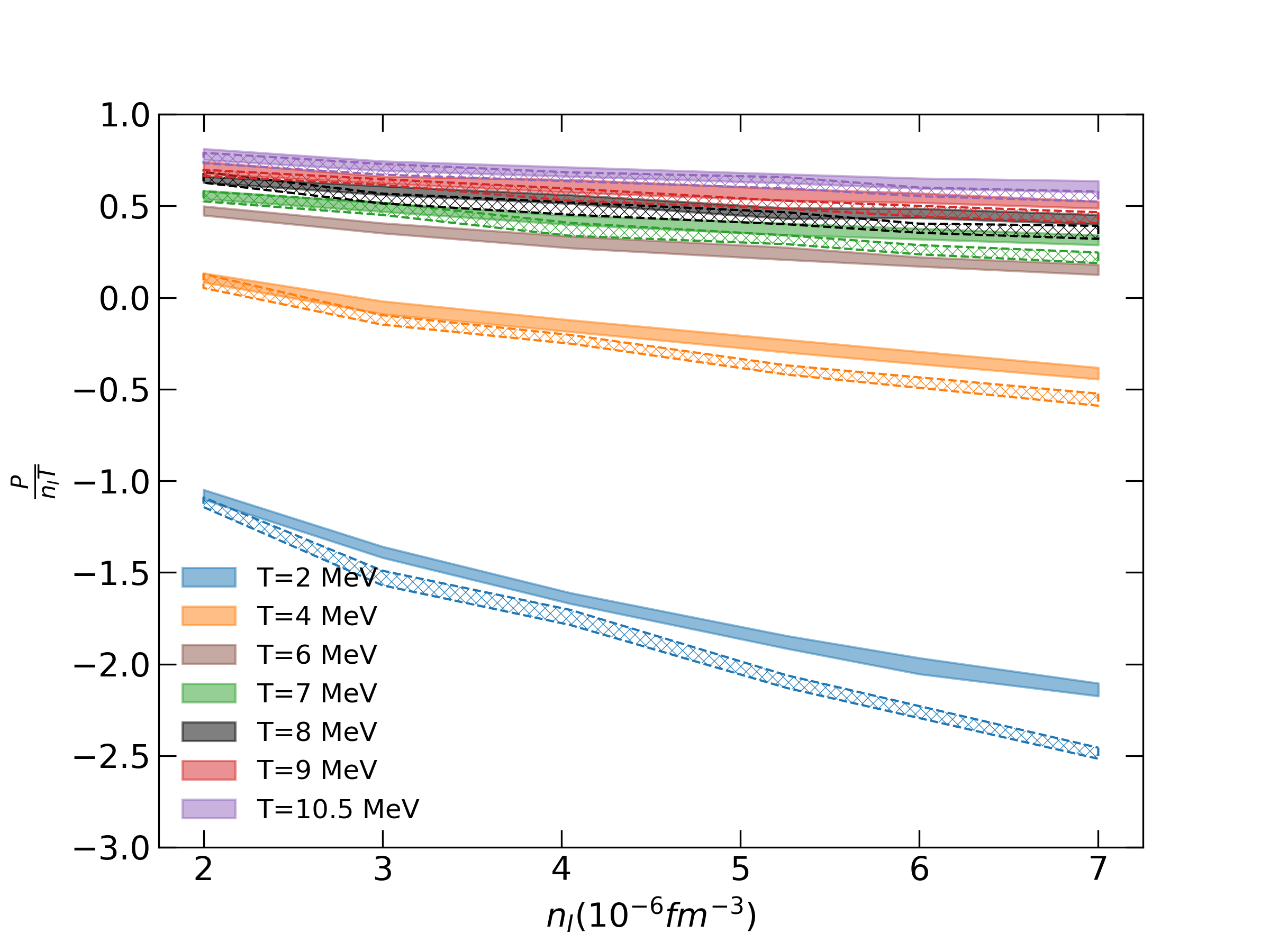}
\caption{Reduced pressure, $P/n_I T\equiv P/P_{id}$, as obtained from our  MD simulations with screened Coulomb, Ewald sums and finite ion width, as a function of ionic density for different temperatures. We use OCP with (Z,A) = (38,128) for a range of temperatures and densities corresponding to $\Gamma_C\in \left[3.84,32.06\right]$. The screening is set at $\kappa=0.622$. At low densities and high T ideal (expanding) behavior is recovered approaching unity, while the pressure reverses sign and becomes attractive for lower T. Solid color bands indicate gaussian ions, whereas dashed ones depict point-like particles. Band width represent standard deviation uncertainty from our MD simulations.}
\label{fig:vEoS}
\end{figure}
In some works where interactions are relatively weak, a comparison is performed with respect to the ideal gas case internal energies and pressure, $U_{id},P_{id}$, see \cite{shahzad}. Therefore real plasma thermodynamic properties are expressed in terms of reduced values i.e. $U/U_{id}$ and $P/P_{id}$, so that typically a departure from unity shows the degree of changes with respect to ideal fluid.

The virial EoS as laid out in Eq. \eqref{virial} represents the limit in which the degrees of freedom of the system are point-like particles. In our previous work \citep{barba1} we showed the effect of introducing a gaussian ion charge spread in the crystal lattice formation. It is also expected that the EoS, and in turn the virial coefficients are tuned by this effect. The finite size of the particles is usually introduced by means of an excluded volume term  $(V-V_{exc})$ in the Van der Waals approach. 

In the low density and non-relativistic ion dynamics involved in this scenario, we seek to model the dependence on the finite width of the $(Z,A)$ nucleus with charge spread $a$ in terms of the adimensional parameter $\eta=1/\sqrt{a}l$. Note $\eta=0$ for point-like particles as the inverse width $a \rightarrow\infty$. Each ion is modeled as a finite-size gaussian charge density distribution as in Eq. \eqref{gauss1} where specifically $a_i=\frac{3}{2\left\langle R^2\right\rangle}$, and following previous work \citep{barba1} we use $\sqrt{\left\langle R^2\right\rangle}=\left(0.8 A^{1 / 3}+2.3\right)\rm fm$, reasonably describing the $A>60$ nuclear sizes.
Thus the in-built excluded volume is proportional to $V_{exc}\sim a^{-3/2}\sim {\left\langle R^2\right\rangle}^{3/2}$ or $\eta \sim (V_{exc}/V)^{1/3}$. We choose to associate each $1/(V-V_{exc}) \rightarrow (1+\eta)/V$ although other parametrizations are indeed possible. In this way we prescribe an expansion for the realistic OCP pressure as

\begin{align}
\frac{P}{k_{\mathrm{B}} T}&=\left(1+\eta\right)n_I+ \left(1+\eta\right)^2 B_{2}^{\prime}(T) n^{2}_I+\left(1+\eta\right)^3 B^{\prime}_{3}(T) n^{3}_I+ \nonumber\\ & \left(1+\eta\right)^4 B^{\prime}_{4}(T) n^{4}_I+\mathcal{O}(n^5_I). \label{eq:virialgauss}
\end{align}

It is easy to see that in the point-like limit, $\eta=0$ and then the standard virial expansion Eq. \eqref{virial} is recovered, that is when $V_{exc}=0$.

In order to illustrate the systems under study in Fig. \eqref{fig:paraview} we have plotted several density isosurfaces for $(Z,A) = (38,128)$ OCP plasmas with average ionic density $n_{I}=1.1 \times 10^{-4} \rm\, fm^{-3}$ and  (right panel) $n_{I}=7 \times 10^{-6} \rm \,fm^{-3}$ together with a continuous color profile indicating local values. We set $k_BT=8$ MeV. The scale $[0,1]$ describes local density variations from 0 to maximum value $n_{I}^{max}$. The plotting software we use is \texttt{Paraview} \citep{AHRENS2005717}. We can see how in the more sparse system the weaker interactions will provide a closer behaviour to ideal, however it is expected that high order correlations due to density-dependent corrections will not be negligible. The depicted two crust ion densities in the OCP approximation used differ roughly an order of magnitude. There is a qualitative, visual, difference in the density profiles: whereas in the smaller density there are bigger voids and high density regions and in denser sample matter is more compactified in smaller clusters. It is important to remember that the MD simulation degrees of freedom are ions, and eventually as it cools down it approaches the crystallized phase, fulfilled at higher densities/lower temperatures than the ones shown.

In Fig.  \eqref{fig:vEoS} we show the EoS for both point-like and realistic ions from our MD simulations, as described in Section \eqref{intro}, i.e. reduced ionic pressure versus density for a range of temperatures and densities corresponding to $\Gamma_C\in \left[3.84,32.06\right]$. The screening is set at $\kappa=0.622$. It is seen that the pressure tends to the ideal gas value when the ionic density is reduced, as expected from Eq.  \eqref{virial}, for each set temperature. The smaller the temperature the lower the pressure goes, even as far as becoming negative for $k_BT \lesssim 4$ MeV. This behaviour has already been reported for less refined  Yukawa OCP not including Ewald summation and gaussian particles, see, for example, \cite{Khrapak2015} and \cite{shahzad}.

\begin{figure}
\includegraphics[width=\linewidth]{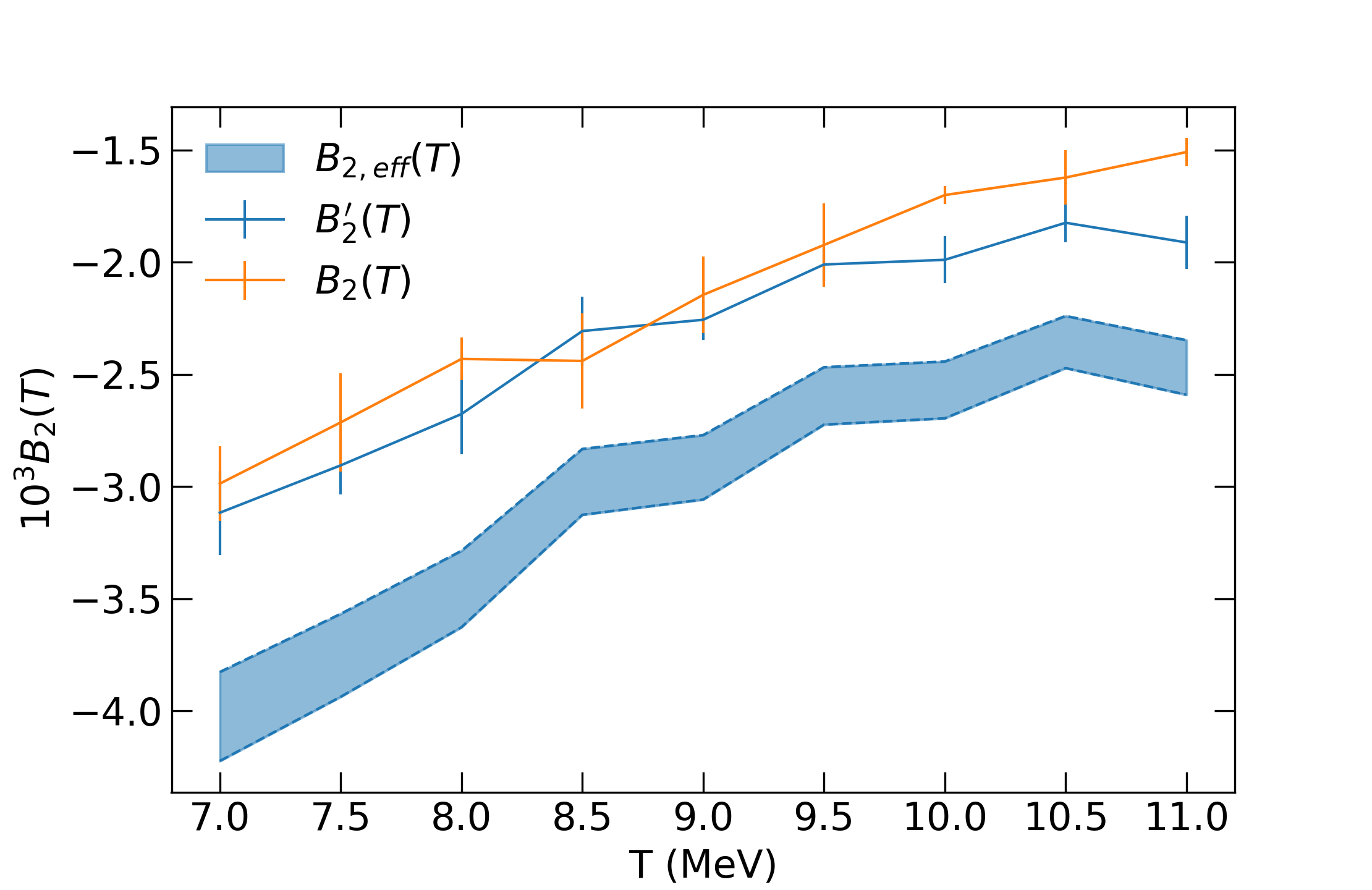}
\caption{Virial coefficient $B_2, B_2^{\prime}$ and $B_{2,\rm eff}$ as a function of temperature from Eq. \eqref{eq:virialgauss}, as extracted from  pressure gas isotherms with $k_BT\geq 7$ MeV. Dependence on finite width ion charge spread and higher order density correlations shift the resulting virial coefficients displaying an effective value $B_{2,\rm eff}$. $\Gamma_C\in \left[3.84,9.16\right]$ and $\kappa=0.622$.}
\label{fig:b2}
\end{figure}

\begin{figure}
\includegraphics[width=\linewidth]{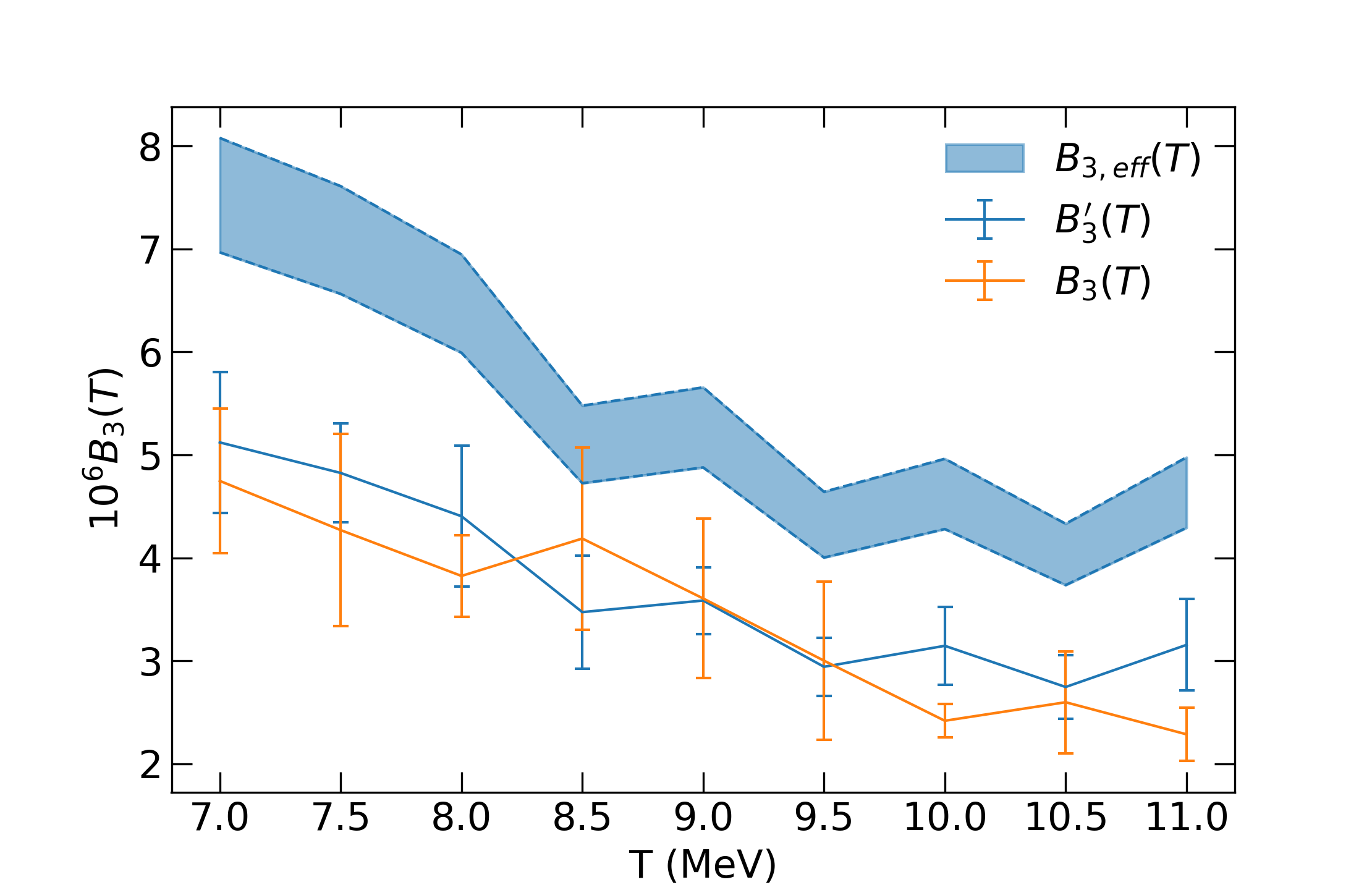}
\caption{ Same as Fig. \eqref{fig:b2} for virial coefficient $B_3, B_3^{\prime}$ and $B_{3, \rm eff}$ }
\label{fig:b3}
\end{figure}

\begin{figure}
\includegraphics[width=\linewidth]{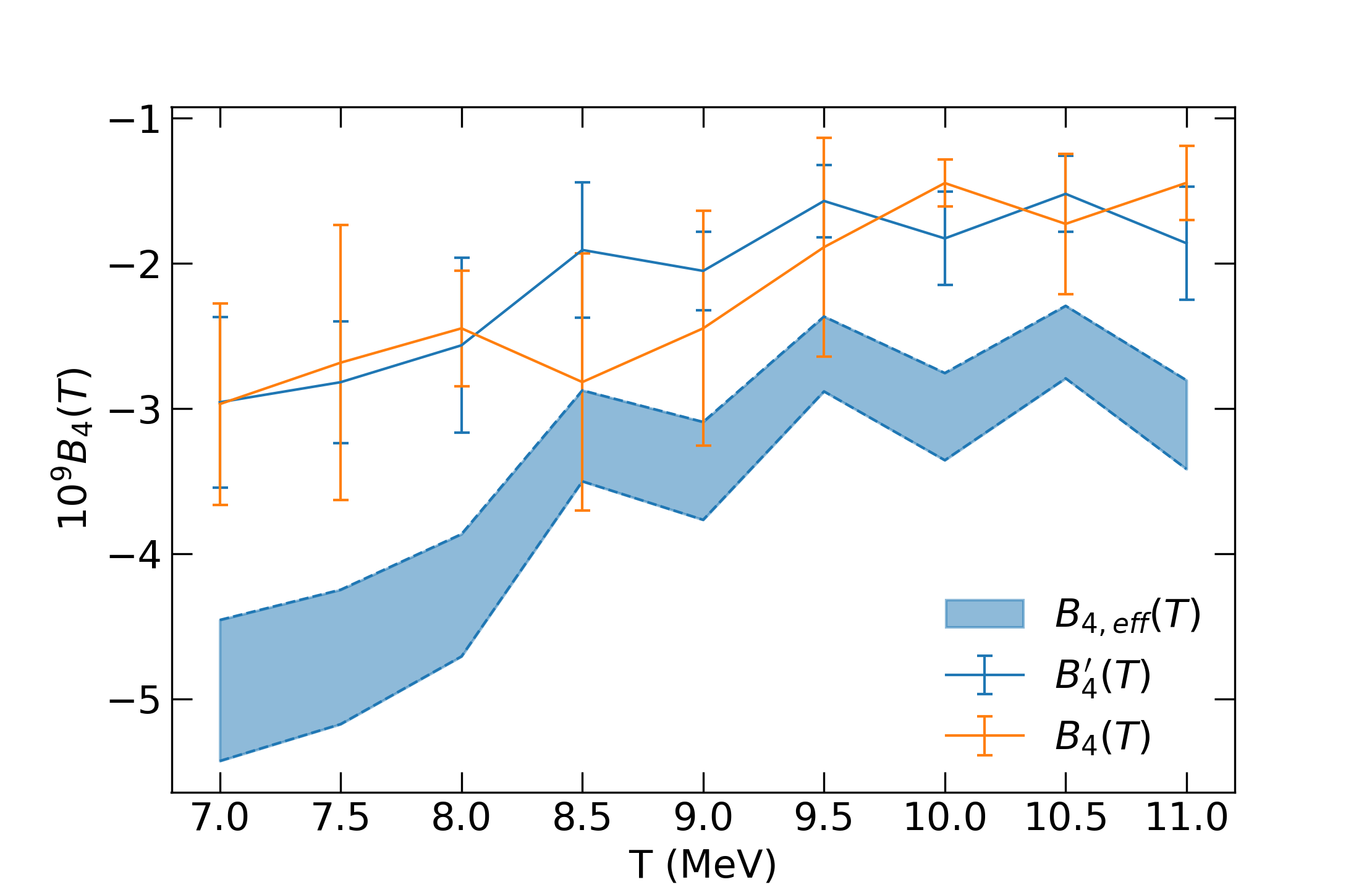}
\caption{Same as Fig. \eqref{fig:b2} for virial coefficient  $B_4, B_4^{\prime}$ and $B_{4,\rm eff}$. }
\label{fig:b4}
\end{figure}

From these curves in Fig. \eqref{fig:vEoS} we fit the virial expansion to our OCP data with ion (Z,A) = (38, 128) using least-squares and obtain the virial coefficients from either Eq. \eqref{virial} or Eq. \eqref{eq:virialgauss} for point-like or gaussian ion runs. All the fits converge and are adequate for the representation of the numerical outputs within the gas regime, when the liquid-gas transition has already ocurred,  in the temperature range $k_B T\in\left[7,11\right] \mathrm{MeV}$ ($\Gamma_C\in \left[3.84,9.16\right])$. The screening is set at $\kappa=0.622$. We show in
Figs. \eqref{fig:b2}, \eqref{fig:b3}, \eqref{fig:b4}  the virial coefficients obtained by this procedure, 
as a function of temperature. In the figure, coefficients $B_k$ have been conveniently normalized by  multiplying them by $\frac{1}{V^{k-1}}$ where $V$ is a typical volume for our simulation box. It is seen that, as expected for the screened Thomas-Fermi interaction, two-body forces are dominant and so from Eqs. \eqref{b2}, \eqref{b3}, $|B_2|>|B_3|>|B_4|$ for the full range of temperatures. This is also consistent with coefficients of increasing order $k$ decreasing  accordingly to add up into the convergent series in Eq. \eqref{eq:virialgauss}. In physical terms, this decline in the coefficients can be better understood when looking at the density profiles of the resultant fluid system with fixed (Z, A) evolved by our MD code, see Fig. \eqref{fig:paraview}.

The effect of the ionic spread and density correlations is shown in Figs. \eqref{fig:b2}, \eqref{fig:b3} and \eqref{fig:b4} for $B_2, B_3, B_4$, respectively. We label the virial coefficients $B^{\prime}_k (T)$ for the gaussian case and $B_k(T)$ for the point-like one (same expression as in Eq.\eqref{eq:virialgauss} when $\eta=0$).

They are perfectly compatible within statistical uncertainty for most of the temperature range considered. Together with them, we consider an effective virial coefficient that we define as $B_{k,\rm eff} = (1+\eta)^kB_k^{\prime}$ and plot it in shaded blue color band for all densities considered. The figures clearly show that density correlations along with finite ion size effects are included in the $\eta$ parameter and play a key role so that the virial expansion  coefficients get clearly shifted as interactions are smeared out when ion charge is spatially distributed. This fact, combined with the aforementioned compatibility of the coefficients, means that the complete information about the finite ionic spread is encoded in the factors $(1+\eta)^k$ that arise in each term of the virial expansion. This, in turn, supports the choice of parametrization in Eq. \eqref{eq:virialgauss}. Thus we not only show the existence of a measurable effect of the more realistic ions, but also are able to compute these effects in terms of the OCP virial expansion excluded volume factors encoded in $\eta$ that can distort point-like values $\sim 25\%$ to $100\%$. We emphasize that a fully consistent MCP system will yield a further $Z,A$ dependence confirming a robust effect already presented in \cite{barba1} along the whole range of densities of interest. Even a taylored A-dependent ion OCP system arising from energy density functionals  will reveal interesting features   regarding the combined effect of heavy species and composition \citep{barba_eos}. This is  typically overlooked in current works, mostly restricting to nuclei with $A\lesssim 8$. In line with this, even the presence of exotic species like the tetraneutron have been shown to largely distort the neutron rich ion fractions, and thus the EoS,  see \cite{paisrefId0}.

\section{ Liquid-gas phase transition and heat capacities in the virialized EoS}
\label{section3}
\begin{figure}
    \centering
    \includegraphics[width=0.5\textwidth]{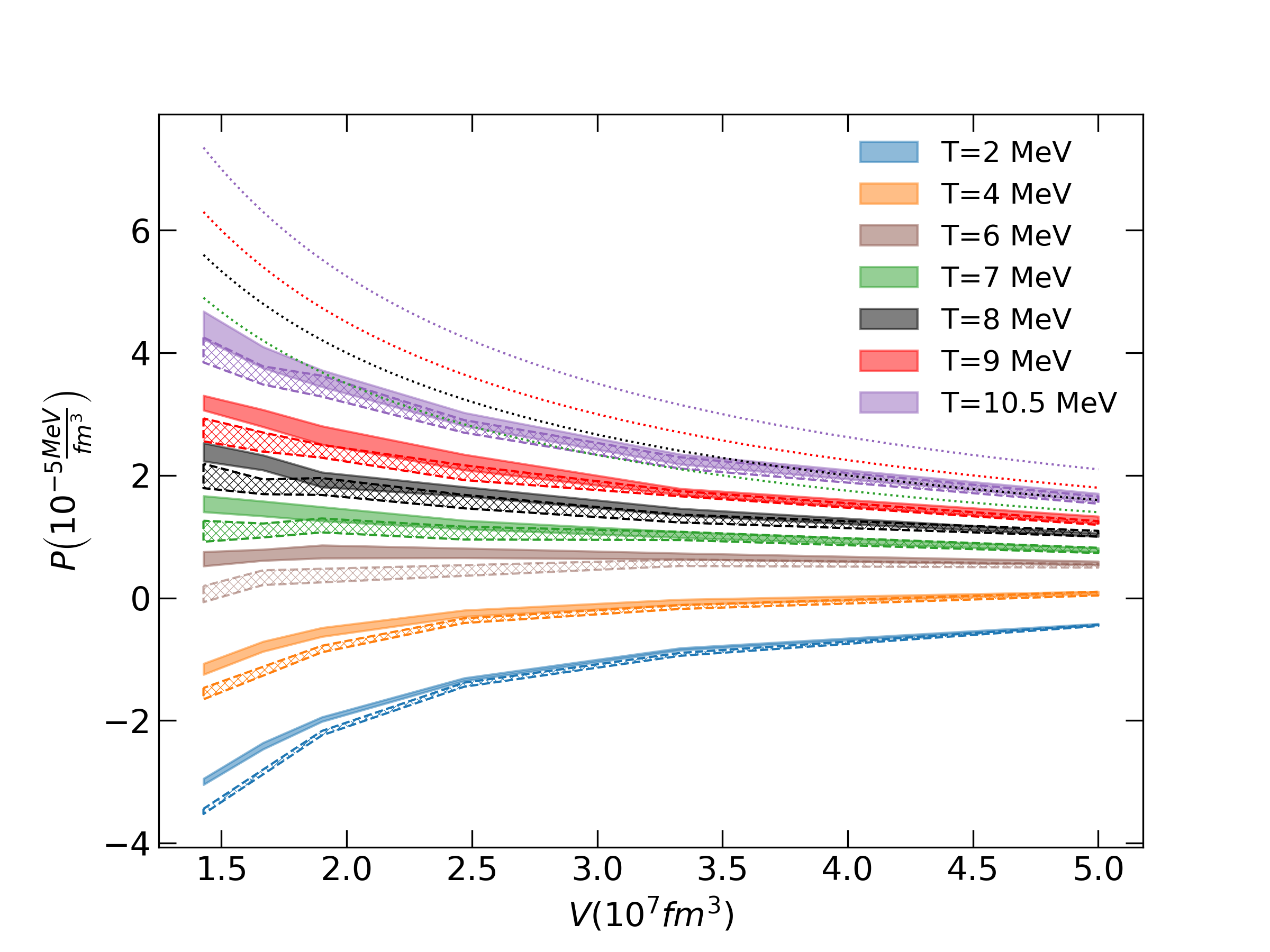}
    \caption{Pressure as a function of volume for OCP samples at different temperatures. We use ions $(Z,A)=(38,128)$. Data are the same as in Fig. \ref{fig:vEoS} with $\Gamma_C\in \left[3.84,32.06\right]$ and $\kappa=0.622$. Ranges reflect uncertainty intervals corresponding to one standard deviation on the pressure. Solid and dashed ranges represent gaussian and point-like ions, while dotted lines are the ideal gas values $P\propto\frac{1}{V}$.One can see a sign change in $\left(\frac{\partial P}{\partial V}\right)_T$ as a function of temperature, going from positive to negative derivative as $T$ grows.}
    \label{fig:PvV}
\end{figure}

The low density screened ion system we simulate  in the range of temperatures and densities $k_B T\in\left[2,11\right] \mathrm{MeV}$, $n_I\in [2,7]\times 10^{-6}$ $\rm fm^{-3}$ can be characterized by the isotherm pressure curves. This is equivalent to a range of temperatures and densities corresponding to $\Gamma_C\in \left[3.84,32.06\right]$. The screening is set at $\kappa=0.622$. In Fig. \eqref{fig:PvV} we plot the pressure as a function of simulation volumes using realistic gaussian charge distribution (solid) and point-like (dashed) ions. 

Our set of MD simulations is used to perform an analysis of the characteristic features of the liquid-gas transition. Due to the non-linear aspects present we choose the powerful techniques of  Artificial Neural Networks (ANN), see appendix \ref{appendix2}. In brief, they consist in a set of layers i.e. input layer, hidden layers and output layer. The hidden layer can be more than one in number, thus referring to deep ANN with each layer consisting of $n$ number of neurons. Each layer will be having an activation function associated with each of the neurons. The activation function is the function that is responsible for introducing non-linearity in the relationship, typically a sigmoid $\sigma(x)$. In our case, the output layer must contain a linear activation function as the output (pressure) has in principle no bounds. Each layer can also have regularizers associated with it. Regularizers are responsible for preventing overfitting. ANN consist of, first, a  forward propagation phase i.e. is the process of multiplying weights with each data entry and adding them to the  bias. Later, a backward propagation phase  updates the weights in the model. Backward propagation requires an optimization function and a loss function, to be selected by the actual implementation of the ANN. 

In appendix \ref{appendix2} we provide an optimized parametrization with two perceptron layers, each with two neurons capable of reproducing the non-linearity of the realistic warm ion gas virialized EoS with improved Ewald sums. 
The explicit form reads

\begin{equation}
    P = \left(0.1407320023\times K_2 (\tilde{n}_{\mathrm{I}}, \tilde{T}) + 10.61859989\right)\times10^{-6} \mathrm{\frac{MeV}{fm^3}},
\end{equation}

where  $K_2 (\tilde{n}_{\mathrm{I}}, \tilde{T})$ is the output of the second perceptron layer depending on reduced input data $\tilde{n}_{\mathrm{I}}, \tilde{T}$. This parametrization produces a relative error of 1.91 \% when predicting pressures within the sample provided in this work. Note that in this example illustrating the OCP results we have kept fixed the ion $(Z, A)$ but without loss of generality this information can be included when constructing and training the ANN. This constitutes a first step and proof of concept of the powerful techniques of machine learning in this context. A more refined version for the full crust will be provided  in \cite{barba_eos}. 

In Fig. \eqref{fig:vEoS} the difference between the ideal  case (dotted lines) and the more realistic case was already depicted under the form of reduced pressure values. From Fig. \eqref{fig:PvV}, we can see the pressure for the spread out charges is always bigger, as the system is less bound. Note this is consistent with our previous results \citep{barba1} regarding the crystal lattices for realistic ions in OCP or MCP mixtures. Equally consistent is the fact that for bigger volumes (lower densities) the finite size effect has a reduced impact, justifying the $\eta$ parametrization used for the virial expansion in Eq. \eqref{virial} developed above.

We observe there is a clear sign change in the derivatives $\left(\frac{dP}{dV}\right)_T$ that will characterize the thermodynamics as the system expands or contracts, that is a projection of the vEoS over the $P-V$ plane. Note that in our setting the Gibbs energy is   only dependent on $V,T$ as the number of particles remains fixed for each simulation (no species conversion) and no external work is exerted on the box walls.
In the pressure isotherms depicted we can clearly differentiate three regions, according to a {\it critical} temperature $T_c$, usually coined as subcritical $(T<T_c)$, critical $(T\sim T_c)$ and supercritical $(T>T_c)$. Regarding the OCP the Maxwell construction determines the coexistence region of liquid and gas, and the critical isotherm with $T=T_c$ that in our setting is located $T_c \simeq 6.5$ MeV. Although in our OCP we do not consider variability in the ion species we expect some  dependence on the charge spread (ion composition).

The thermal properties of the NVT system undergoing a phase transition are better understood when looking at the heat capacities, i.e. the thermodynamical potential response to energy variations with temperature.
Heat capacities are defined, generically, as

\begin{equation}
    C_{V}=\left(\frac{\partial \rho }{\partial T}\right)_{V} \quad  C_{P}=\left(\frac{\partial \rho}{\partial T}\right)_{P}
\end{equation}

with $C_P$ and $C_V$ the heat capacities at constant pressure and volume, respectively.

In our screened system the thermal behaviour is determined by the impact of the electron component. This is in-built in the screened potential caused by effective forces driving the inter-particle interaction. Generically the two components that largely dominate the specific heats in the crust behave in a differentiated way.  On the one hand  the specific heat capacities $c_V,c_P$ for the ultra-relativistic electrons $m_e \sim 0$ forming a highly degenerate electron Fermi gas will contribute to the heat capacity per unit volume \citep{potekhin} at $T \ll T_{\mathrm{F},e}$ with $T_{\mathrm{F},e}=E_{F,e}/k_B$ as 
\begin{equation}
c_{\mathrm{V}, e}=n_e \frac{\pi^2}{k_{\mathrm{F}, e} c} T .
\end{equation}
It is usually assumed that in the outer crust the main contribution is given by that of the ions since electrons are highly degenerate and $C_{e,V}\approx C_{e,P}$. In an ideal classical crystal, the ion heat capacity is $C_{\mathrm{I}}=3 k_{\mathrm{B}}$ while for ideal gas $\sim 3/2 k_B$. Quantum effects strongly reduce, however,  this value for temperatures much lower than that associated to ion plasma frequency that we will not consider in our scenario.

\begin{figure}
    \centering
    \includegraphics[width=0.5\textwidth]{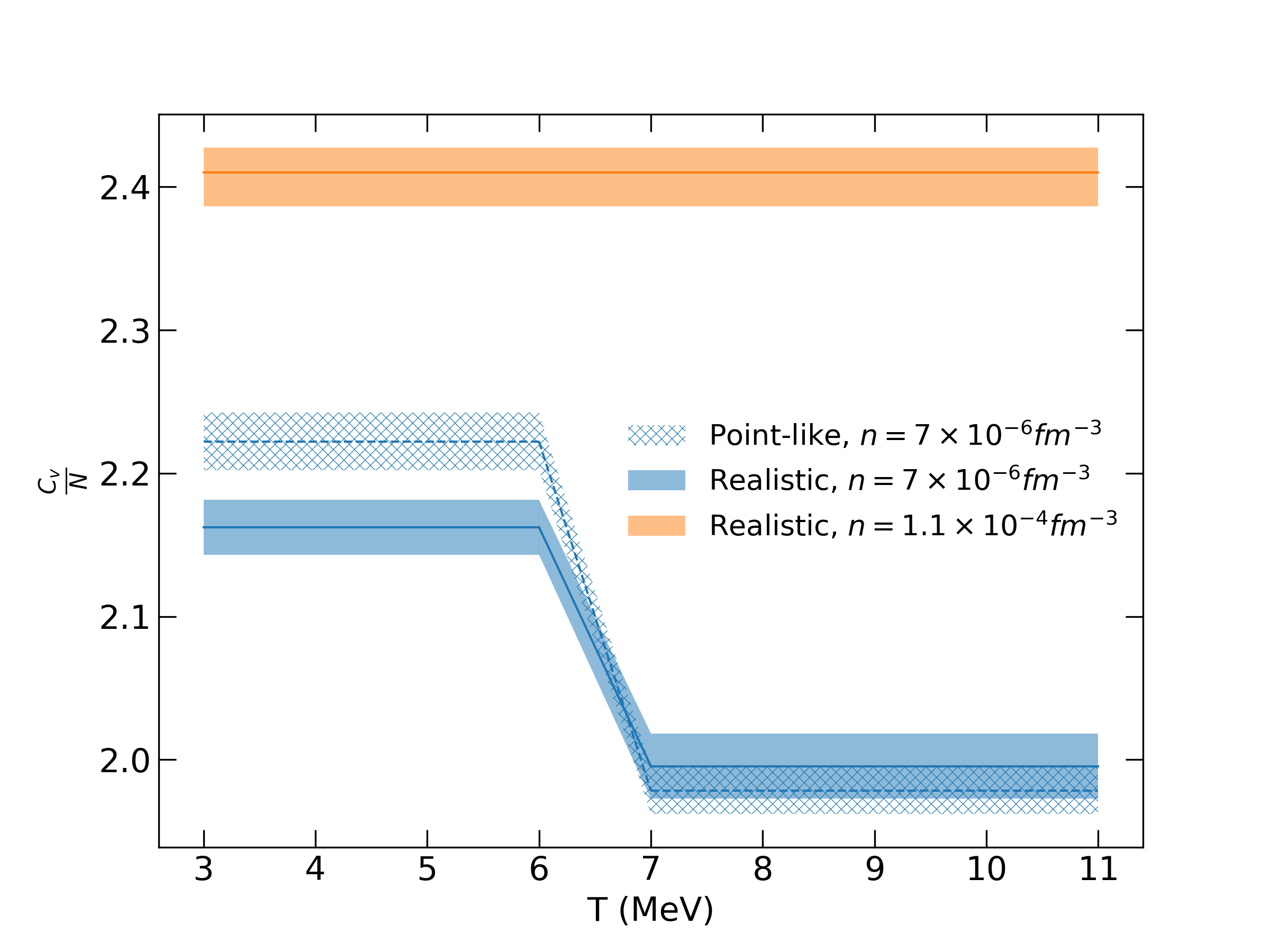}
    \caption{Specific heat per ion at constant volume, $\frac{C_V}{N}$, for screened interactions using Ewald sum at densities $n_I= 7\times 10^{-6}$ (blue lines, $\Gamma_C\in \left[5.82,21.37\right]$) and $n_I= 1.1\times 10^{-4}$ (orange line, $\Gamma_C\in \left[14.60,53.53\right]$)for a common screening parameter $\kappa=0.622$. For the lower density case there is a discontinuity signaling the onset of the liquid-gas phase transition while this happens at higher $T$ for the $n_I= 1.1\times 10^{-4}$ case, thus not shown. Realistic ion sizes (solid lines) shift $C_V$ per ion as compared to point-like (dashed lines) ions. We include associated error bands corresponding to our simulations and fitting procedure, see text for details.}
    \label{fig:cvN200}
\end{figure}

As seen from the $P$ isotherms shown in Fig. \eqref{fig:PvV} the sign change in the  derivative  $\left(\frac{dP}{dV}\right)_T$ marks the onset of a 1st-order transition.
Namely for our ion species the critical isotherm lies approximately $T_c \simeq 6.5$ MeV, and has an associated flat profile of pressure as a function of V, i.e. $\left(\frac{dP}{dV}\right)_T=0$. We also note that below and above the critical isotherm the ion system has a discontinuity  $C_{V}=\left(\frac{\partial \rho }{\partial T}\right)_{V}$ as depicted in Fig. \eqref{fig:cvN200} for an illustrative selected ion density $n_I= 7\times 10^{-6}$ $ \rm fm^{-3}$ (blue lines). We have corroborated this by fitting for the slope of the $\rho(T)$ data in the two different temperature ranges, before and after the expected first order phase transition. We include associated error bands corresponding to the fitting procedure from our statistically distributed simulation energy values. For the sake of comparison we also plot the case with $n_I= 1.1\times 10^{-4}$ $\rm fm^{-3}$ (orange line) as it does not undergo any phase transition at this temperature range. We note that in both cases the value is largely increased $\sim 40\%$ with respect to the ideal gas value $\sim 3/2 k_B$ in a non-magnetized system. Realistic ion sizes (solid lines) shift $C_V$ per ion as compared to point-like (dashed lines) ions less than $\sim 5\%$.

In addition to the ion heat capacity at constant volume we have also studied the ion heat capacity at constante pressure. Note that since we simulate our ion system under the NVT ensemble (where pressure is not longer constant) we calculate this quantity indirectly as deduced from the generalized Mayer relation $C_P-C_V$, a well known relation between the isochoric and isobaric heat capacities. The explicit form is 

\begin{equation}
    C_P-C_V = -T\frac{\left[\left(\frac{\partial P}{\partial T}\right)_V\right]^2}{\left(\frac{\partial P}{\partial V}\right)_T}
    \
    \label{eq:mayer}
\end{equation}
In the above equation $\left(\frac{\partial P}{\partial T}\right)_V$ is the inverse of the thermal expansion coefficient, and $\left(\frac{\partial P}{\partial V}\right)_T$ is $\frac{1}{V}$ times the inverse of the isothermal compressibility. The system that we are simulating, the cooling crust of a proto-NS or warm Supernova matter, is a non-ideal gaseous-liquid system, which at very high temperature is expected to reproduce the ideal gas Mayer's relationship. This is illustrated by substituting the virial expansion Eq. \eqref{eq:virialgauss} into Eq. \eqref{eq:mayer}. In this way for the realistic vEoS with gaussian ion corrections we obtain

\begin{equation}
 \frac{C_P-C_V}{N} = \frac{\left[\left( \sum_{k=1}^{\infty} n_I^k (1+\eta)^k \frac{\partial B^{\prime}_k}{\partial T}\right)\right]^2} { \left( n_I (1+\frac{4\eta}{3}) \sum_{k=1}^{\infty} n_I^k (1+\eta)^{k-1} B^{\prime}_k \right)}
 \end{equation}

\noindent where we define $B^{\prime}_1=1$. Note that for a point-like approximation in a non-interacting system we recover the usual ideal Mayer relation $(C_P-C_V)/N =1$.

From the $P-V$ behaviour in Fig. \eqref{fig:PvV} of the critical isotherm limiting  with associated vanishing of the derivative $\left(\frac{dP}{dV}\right)_T=0$, together with the fact that the derivative $\left(\frac{dP}{dT}\right)_V$  is finite, means that the difference between $C_P$ and $C_V$ from Eq. \eqref{eq:mayer} diverges. The conclusion that it is $C_P$ which does diverge can be arrived to by means of realizing that $\frac{E}{N}$ vs $T$ curves are smooth for any of the densities in the studied range, thus $C_V$ always remains a finite value. This further confirms the phase transition as first-order. 

\begin{figure}
    \centering
    \includegraphics[width=\columnwidth]{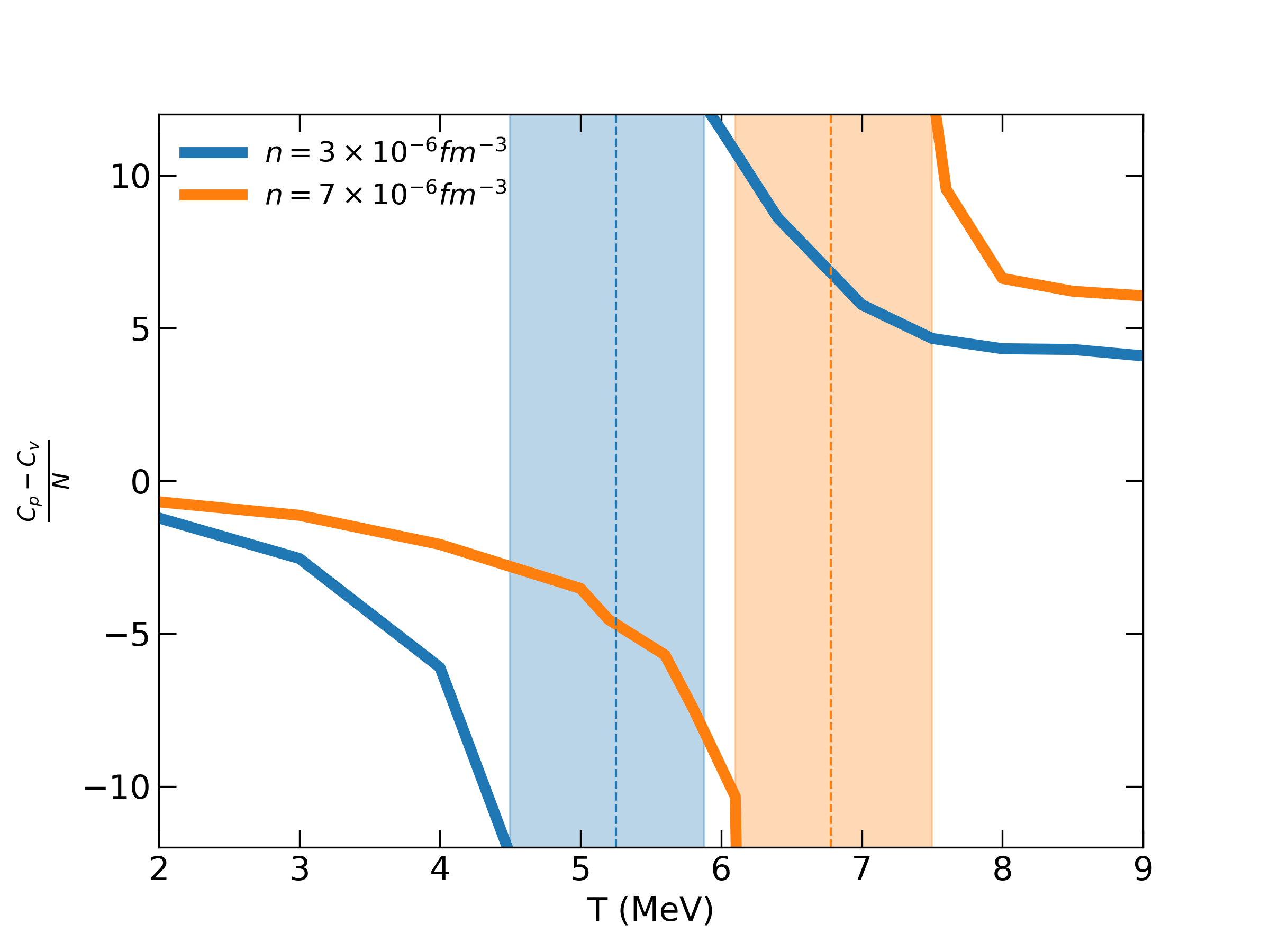}
    \caption{Ion Mayer relation $\frac{C_P-C_V}{N}$ as a function of temperature for  densities $n_I= 3\times 10^{-6}$ (blue line) with $\Gamma_C\in\left[5.37,24.17\right]$ and $n_I= 7\times 10^{-6}$ (orange line) corresponding to $\Gamma_C\in\left[7.12,32.06\right]$ for finite size ion species $(Z, A)= (38, 128)$  corresponding to $\kappa=0.622$  The metastable region with liquid-gas coexistence is depicted to the left of each asymptote (dashed vertical line in the shaded regions).}
    \label{fig:mayer}
\end{figure}

In Fig. \eqref{fig:mayer} we show the Mayer's relation for two densities $n_I= 3\times 10^{-6}$ $\rm fm^{-3}$ (blue line) and densities $n_I= 7\times 10^{-6}$ $\rm fm^{-3}$(orange line) for finite size ion species $(Z, A)=( 38, 128)$. We can clearly see that divergence signaling the liquid-gas phase transition appearing at two different temperatures (larger as density grows) corresponding to $\Gamma_C\sim 10$ for both densities. The metastable region with phase coexistence is depicted to the left of each asymptote (dashed vertical line in the shaded regions). The negative values $C_P-C_V$ must be understood as the presence of a coexisting liquid-gas phase.
Note that in the finite temperature ion gas phase $C_P/N \sim 8$ being a factor nearly 4 larger than its isochoric counterpart $C_P\sim 4 C_V$. This fact may further impact the cooling behaviour of species in gas phases in the warm matter later forming the outer NS crust as derived from the heat equation Eq. \eqref{heat}.

In order to compare our findings in the  OCP setting used in this work to other different warm EoS in the literature incorporating clusters, we have selected a canonical example of the Relativistic Mean Field (RMF) model  calculation i.e. TM1e \citep{2020ApJ...891..148S}  including clusters up to $A=8$. It is important to emphasize that in our scenario we deal with OCP systems, so that pressure and energy density are generated  by a single species while in the TM1e this is done by a few of them arising from thermodynamic conditions previously settled. Typically, the heavy species formed near the liquid-gas boundary for Supernova matter, as presented in  \cite{Ishizuka_2003}, are usually not considered in this scenario as the lepton rich matter (and later on neutrino-less matter) cools down. Our scenario agrees with that presented by Ishikuza et al as our transition arises at about $\sim 1.4T_{\rm boil}$ their boiling temperature at the same densities but computed in a very  different framework. In these RMF  models, due to the light nuclei involved in the calculations, see also \cite{paisrefId0} in the same fashion, we expect that pressure obtained will be systematically higher. In the work of \cite{murarka2022} using finite ion degrees of freedom in the OCP scenario, the NS outer crust EoS at $T=0$ was constructed based on the predictions of a Machine Learning algorithm from AME2016 nuclear dataset immerse in a relativistic electron gas and lattice dynamics. As a NS cools down, a possible MCP distribution becomes more peaked so that at  cristallization one can consider that the composition is frozen and unchanged to the $T=0$ ground state. In this spirit we have used the composition for the NS outer crust computing  warm EoS values in the density-temperature range we study. For the sake of discussion, we considered illustrative conditions  near the liquid-gas phases at $T=10$ MeV and neutron rich clusterized matter at $Y_p\sim0.3$ were attained. For an ion  density $n_I=2 \times10^{-6}\, ( n_B=2.56\times 10^{-4}$  $\rm fm^{-3})$ which gives a Coulomb parameter $\Gamma_C=4.22$ our OCP simulation yields $P_{0,\rm OCP} = 1.0672\times 10^{-5}$ $\rm MeV\,fm^{-3}$ while $\epsilon_{0,\rm OCP} =0.240310$ $\rm {MeV}{fm^{-3}}$.
In the RMF value pressure is a factor $\sim 300$ larger i.e. $P_{\rm Shen}\sim 300 P_{0,\rm OCP}$ with a similar value of $\epsilon_{\rm Shen}\sim \epsilon_{0,\rm OCP}$. Instead for the cold OCP approach with degenerate electrons deduced by \cite{murarka2022} $P_{\rm Murarka}\sim 50 P_{0,\rm OCP}$ and similar value for $\epsilon$ so that $\epsilon_{\rm Murarka}\sim \epsilon_{0,\rm OCP}$. In this last case the ion predicted is $(Z, A)=(36, 116)$ at the fixed density value. Both warm RMF and cold NS outer crust EoS yield systematically larger pressure values when compared to our simulated samples. 
Thus we conclude the finite size ion degrees of freedom can not be overlooked in the SN matter or proto NS involving heavy ions as the full picture can not be understood without them. The determination of the liquid-gas transition is key to the conditions of heavy cluster formation, and temperatures up to a dozen MeV will play a role, thus having an impact on the outer layers of cooling proto NS.

\section{Conclusions}
\label{section4}
We have performed microscopic Molecular Dynamics simulations with periodic boundary conditions of warm and neutral plasmas with spin-zero ions interacting by electron screened Coulomb potentials at finite temperature. We have used efficient Ewald sums for finite size gaussian ion charge distributions in the system. We have focused on the OCP description with single ion species characteristic of a given density in the crust. These conditions are of interest to describe low density Supernova matter and early phases of proto neutron stars. Although more refined treatments involving MCP systems are indeed possible we study and characterize this type of matter and the liquid-gas phase transition it undergoes and find distinctive features with our improved treatment. We compute numerical expressions for the energy density and stress tensor for screened gaussian ion charges including the additional screening $\alpha_{\rm Ewald}$ introduced by the Ewald summation method for the first time. By using this method low density thermodynamical quantities are accurately described to longer distances. In addition our system also displays a typical Thomas-Fermi screening length from the relativistic degenerate electron plasma that we effectively take at $T=0$.

We dynamically follow the positions, velocities and accelerations from the prescribed set of conditions in a NVT ensemble that we thermalize before results are obtained in a temperature range where crystallization does not happen. For the fluid phases we describe a set of pressure isotherm curves $P_T(V)$ as a function of volume where metastable coexisting (liquid-gas) and stable (gas) phases appear as T grows. 
 We are able to determine differences among the point-like and finite size ion treatments for the selected species so that when parametrizing the gas EoS clear differences appear in the virial coefficients up to 4th order, the meaningful range where we restrict our calculations. For the systems we explored these corrections to $B_k$ can be as large as $\sim 25-100\%$. In particular we extract density and ion size dependence (by means of an extra parameter $\eta$) in terms of a $1+\eta$ power expansion, so that at low densities we consider the excluded volume effects. In the usual treatments this parameter $\eta=0$ as ions are considered point-like. In our OCP calculation we also find dependence of the ion mass although we can not further size its dependence leaving this for a future contribution.  When comparing our findings the warm OCP yields systematically smaller pressures than RMF or clusterized matter using the same composition predicted for $T=0$ NS outer crust. Although a more refined MCP system will improve this picture we expect this reduction is robust and due to finite-size ions.

 As an illustrative example of usability we provide an ANN giving a rather efficient expression of the virial EoS in our OCP system $P(T, n_I)$ leaving for a future work a more detailed MCP calculation. As we can capture the dynamical transition liquid-gas in the MD treatment we are able to determine the energy and pressure derivatives involved in the heat capacity at constant volume and constant pressure by generalizing the Mayer relation for the Yukawa potential system with finite size particles for the first time. We find the ion heat capacity $C_V$ is no longer T dependent with its value being $\sim 40\%$ larger than the usual non-magnetized ideal gas value $\sim 3/2k_B$. We can clearly see that the discontinous jump in the $C_V$ signals the liquid-gas transition critical temperature in our simulated system. This confirms the first-order phase transition. From this we are able to determine $C_P$ using the generalized Mayer relation accessible from data in our simulation. We find both ion heat capacities are different in the gaseous phase so that cooling behaviour of systems kept under constant V or P would be different. We foresee that with a MCP or OCP parametrization of crust densities Supernova matter or  proto-NS early cooling phases could be affected under the usual paradigm of equivalent idealized ion $C_V, C_P$ values.

This calculation constitutes an example of the complementary study of microscopic models based on energy functionals where particle dynamics can provide the effective description of screened ions with improved force and energy summed techniques. Early phases of Supernova matter where a distribution of nuclei with some heavy species and low mass nuclei are present can benefit from our virialized EoS at finite temperature. Preliminary comparisons with RMF and energy functional minimization from AME2016 datasets show a distintive feature, with those EoS pressure values over the ones estimated in the OCP scenario even is the liquid-gas boundary yields similar boiling temperatures. Composition is thus playing an important role and heavy species with $A>50$ must be present in the consistent picture when including the proto NS. Note also that under dissolution or charge-changing reactions this description must incorporate different species as well. Even if OCP is most surely not realized in the polluted scenarios in the aftermath of a Supernova explossion we foresee our results remain valid for each species component with finite size ion charge $(Z,A)$ at given density during its associated lifetime.  
\section*{Acknowledgments}

We acknowledge  useful comments from  A. Aguado, N. Chamel, P. Char and A. Fantina. This work has been supported by Junta de Castilla y León SA096P20
and Spanish MICIN grant PID2019-107778GB-I00, RED2022-134411-T and PID2022-
137887NB-I00. D.B.G. acknowledges support from a Ph.D. Fellowship funded by Consejería de Educación de la Junta de Castilla y León and European Social Fund. Computation in this project has been performed with TITAN, DRACO multi-core machines at University of Salamanca, and Supercomputing centers SCAYLE and CETA-Ciemat and Spanish RES resources.

\section*{Data Availability}

 The computed data presented and discussed in this paper will be shared upon reasonable request.



\bibliographystyle{mnras}
\bibliography{apssamp} 




\appendix
\section{Energy and force for electron screened Coulomb potentials with  gaussian spread ion charges}
\label{appendix1}

In this section we detail the contributions to the potential energies and forces for a system of electron screened gaussian ions using the Ewald sum procedure. In brief in this method real ion charges are supplemented by spurious compensating charges (screening charges) in order to effectively separate the original interaction into  short and long range parts and a self-interaction contribution as explained in our previous work \citep{barba1}.
Each ion is modeled as a finite-size Gaussian charge density distribution in the form $\rho_{i, a_i}(r)=Z_i\left(\frac{a_i}{\pi}\right)^{\frac{3}{2}} e^{-a_i r^2}$,where specifically $a_i=\frac{3}{2\left\langle R^2\right\rangle}$ is related to the average square $(Z_i,A_i)$ ion radius.

A short-range part describes  the interaction between the real charges and the potential created by the sum of real plus screening charges, 

\begin{equation}
    \phi_{\rm{short-range},i}\left(\vec{r}\right)=    \phi_{Zi,a_i}\left(\vec{r}\right)+    \phi_{-Z_i,\alpha_{\rm Ewald}}\left(\vec{r}\right).
    \label{phishort}
\end{equation}

Here the potentials on the right hand side are obtained by solving the Poisson's equation for a gaussian charge density distribution with total charge $Z_i (-Z_i)$ and square inverse width $a_i (\alpha_{\rm Ewald})$ which induces a Debye-Hückel-type potential

\begin{multline}
    \phi_{Zi,a_i}\left(\vec{r}\right)=\frac{Z_i}{2|\vec{r}-\vec{r_i}|} e^{\frac{1}{4 a_i \lambda_e^2}}\left[e^{-\frac{|\vec{r}-
\vec{r_i}|}{\lambda_e}}\mathrm{erfc}\left(\frac{1}{2\sqrt{a_i}\lambda_e} - \right.\right.\\ 
\left.\left. \sqrt{a_i}|\vec{r}-\vec{r_i}|\right)  -
e^{\frac{|\vec{r}-\vec{r_i}|}{\lambda_e}}\mathrm{erfc}\left(\frac{1}{2\sqrt{a_i}\lambda_e}+\sqrt{a_i}|\vec{r}-
\vec{r_i}|\right) \right],
\end{multline}

and 

\begin{multline}
    \phi_{-Zi,\mathrm{\alpha_{Ewald}}}\left(\vec{r}\right)=\frac{-Z_i}{2|\vec{r}-\vec{r_i}|} e^{\frac{1}{4 \mathrm{\alpha_{Ewald}} \lambda_e^2}}\left[e^{-\frac{|\vec{r}-
\vec{r_i}|}{\lambda_e}}\mathrm{erfc}\left(\frac{1}{2\sqrt{\mathrm{\alpha_{Ewald}}}\lambda_e} - \right.\right.\\ 
\left.\left. \sqrt{\mathrm{\alpha_{Ewald}}}|\vec{r}-\vec{r_i}|\right)  -
e^{\frac{|\vec{r}-\vec{r_i}|}{\lambda_e}}\mathrm{erfc}\left(\frac{1}{2\sqrt{\mathrm{\alpha_{Ewald}}}\lambda_e}+\sqrt{\mathrm{\alpha_{Ewald}}}|\vec{r}-
\vec{r_i}|\right) \right],
\end{multline}

The short-range total potential energy for a system of these charges is given by the integral expression

\begin{multline}
    U_{\mathrm{short-range}}=\frac{1}{2}\sum_{i=1}^{N_I}\sum_{j\neq i=1}^{N_I} 2 Z_j\left(\frac{a_j}{\pi}\right)^{\frac{1}{2}} \frac{e^{-a_j r_{ij}^2}}{r_{ij}} \\ \int_{0}^{\infty} r' \phi_{\rm{short-range},i}\left(r'\right)e^{-a r'^{2}}\mathrm{sinh} \left(2a_jr_{ij}r'\right)dr' \\
 -\frac{2\pi}{V} \sum_{i=1}^{N_I}\sum_{j=1}^{N_I} Z_j \int_0^{\infty} r'^{2}\phi_{\rm{short-range}, \mathrm{i}} \left(r'\right)dr',
 \label{Ushort}
\end{multline}
where $r_{ij}=|\vec{r_i}-\vec{r_j}|$ is the distance between the $ij$ pairs. Note that this expression explicitly allows for the computation of the energy for different ionic species, although we will restrict to OCP in this work. This contribution has a dependence in the interparticle distance that tails off quickly, so that it converges very rapidly in real space. The choice of the Ewald parameter $\alpha_{\rm Ewald}$ is relevant. We follow the prescription given in \cite{2013JChPh.139x4108H} whereas by imposing a controlled tolerance with respect to the exact value of the energy, both $\alpha_{\rm Ewald}$ and the number of simulation box replicas is fixed ($n_{k,max}=6$). This choice is such that the short-range part converges in the main simulation box, and so the minimum image convention can be applied.

A long-range part of the interaction is created by the compensating charges in all periodically generated simulation boxes, and the background average charge,  $\rho_{\text {avg }}=\frac{\sum_{i} Z_{i}}{V}$ under the prescription  $\sum_i   \rho_{Z_i,\alpha_{\rm Ewald}}-\rho_{\text {avg }}$. Explicitly, this is done by transforming Poisson's equation from the coordinate space to the Fourier $k$-space and including a summation over reciprocal lattice vectors so that the sum converges to a finite value under the form

\begin{equation}
\phi_{\rm long-range}(\vec{r})=\sum_{j}^{N} \sum_{\vec{k} \neq 0} \frac{4 \pi Z_{j}}{V\left(k^{2}+\frac{1}{\lambda^{2}_e}\right)} e^{\frac{-k^{2}}{\alpha_{\mathrm{Ewald}}}} e^{i \vec{k}\left(\vec{r}-\overrightarrow{r_{j}}\right)}
\end{equation}

where $\vec{k}=\frac{2 \pi}{L}\left(n_{x}, n_{y}, n_{z}\right)$ and $n_{x}, n_{y}, n_{z} \in \mathbb{Z}$. 

The associated energy term, $U_{\mathrm{long-range}}$, is thus

\begin{multline}
    U_{\mathrm{long-range}}= \frac{1}{2}\sum_{i,j=1}^{N_I} \sum_{\vec{k}\neq 0} \frac{4\pi Z_i Z_j}{V \left(k^2+
\frac{1}{\lambda_e^2}\right)} \times \\
e^{\frac{-k^2}{4}\left(\frac{1}{a_i}+\frac{1}{\alpha_{\mathrm{Ewald}}}\right)} 
e^{i \vec{k}\left(\vec{r_i}-\vec{r_j}\right)}.
\label{Ulong}
 \end{multline}
Finally, it remains to substract the interaction between the real charge and its own compensating charge as it is included spuriously in the long-range part. It is given by

\begin{equation}
    U_{\mathrm{self}}=2\pi\sum_{i=1}^{N_I} \left(\frac{a_i}{\pi}\right)^{\frac{3}{2}}Z_i \int_0^{\infty} r'^{2} \phi_{Z_i,\alpha_{\mathrm{Ewald}}}\left(r'\right) e^{-a_i r'^{2}} dr',
\end{equation}

In our approach using MD simulations the pairwise force computation  $\vec{F}_{\mathrm{ij}}$, used for obtaining the ion accelerations, arises from the gradient of Eqs. \eqref{Ushort}, \eqref{Ulong} and takes the form

\begin{equation}            
    \begin{aligned}
    &\vec{F}_{\mathrm{ij},\text{short-range}} = 2\left(\frac{a_j}{\pi}\right)^{\frac{1}{2}}\frac{Z_j e^{-a_j r_{ij}^2}}{r_{ij}^2}\left(\frac{\vec{r}_{ij}}{r_{ij}}\right)\times\\
    &\left\{\int_0^{\infty} r^{\prime}\phi_{\rm{short-range},\mathrm{i}}\left(r^{\prime}\right)e^{-a_j r^{\prime 2}}\right. \\
    &\bigg.\left[\left(1+2a_j r_{ij}^2\right)\sinh\left(2a_j r_{ij}^2\right)-2a_j r_{ij}r^{\prime}\cosh\left(2a_j r_{ij}^2\right)\right]dr^{\prime}\bigg\},
    \end{aligned}
\end{equation}

while for the long-range part 
\begin{equation}
 \begin{aligned}
    &\vec{F}_{\mathrm{i,long-range}}=\frac{1}{2}\sum_j\sum_{\vec{k}\neq 0}\frac{8 \pi Z_i Z_j}{V \left(k^2 + \frac{1}{\lambda^2}\right)} e^{\frac{-k^2}{4\alpha_{\mathrm{Ewald}}}}\times\\
    &\left[{e^{-\frac{k^2}{4a_i}}e^{i\vec{k}\cdot\left(\vec{r}_i-\vec{r}_j\right)}-e^{-\frac{k^2}{4a_j}}e^{-i\vec{k}\cdot\left(\vec{r}_i-\vec{r}_j\right)}}\right] \frac{\vec{k}}{2i}.
 \end{aligned}
\end{equation}

\section{Multilayer Artificial Neural Network as a tool to predict warm realistic ion EoS}
\label{appendix2}
In this section we provide an expression for pressure as a function of ion density and temperature for our OCP warm ionic system. For this task we perform regression analysis using Artificial Neural Networks (ANN). Although more refined modelization with is under way \citep{barba_eos} it constitutes a proof-of-concept about improving the usability of our results. ANN is a computer system developed on the basis of biological nervous system, it imitates the neural tissue of the brain in the human body to work together, and is a network system composed of a large number of process or units. The ANN must be understood as a non-linear composite function. We use quantities $\omega_{i j}^h$ to represent the weight from the $i$ neuron in the $h$ layer to the $j$ neuron in the $h$ layer. We use here $\theta_j^h$ to represent the displacement of $j$ neurons in the $h$ layer. The output value $O_j^h$ is demonstrates the value on the $h$ layer. The vector $x$ is the input data of neural network. Its main formula is as follows

\begin{equation}
O_j^h(x)=\sigma\left(\sum_j \omega_{i j}^h x_j^{h-1}+\theta_j^h\right) ,
\end{equation}

\noindent with the activation function or sigmoid
\begin{equation}
\sigma\left(z^h\right)=\frac{1}{1+e^{-z^h}} .
\end{equation}

Neural network training consists in setting up weights as well as basis. Here $O_j^h$ are calculated between the exact amount of $z^h=\omega_{i j}^h x^{h-1}+\theta_j^h$. We call this weighted input, then $O_j^h(x)=\sigma\left(z^h\right)$.

In our particular case our input data for the ANN are the values of pressure, $P$, ion density, $n_I$ and temperature, $T$ in the hypercube $T\in [2,11]$ MeV, $n_I\in [2,7]\times 10^{-6}$ $\mathrm{fm^{-3}}$ and $P \in [-0.43,4.83]\times 10^{-5}$ $\mathrm{\frac{MeV}{fm^3}}$.  We use a reduced version of them, $\tilde{n}_{\mathrm{I}}, \tilde{T}$ as they are  easier to handle i.e. 

\begin{equation}
\begin{aligned}
    &\tilde{n}_{\mathrm{I}} = \frac{\frac{n_{\mathrm{I}}}{10^{-6} \,\rm fm ^{-3}}-4.40500021}{1.86334002},
\end{aligned}
\end{equation}

\begin{equation}
\begin{aligned}
    &\tilde{T} = \frac{\frac{T}{\rm MeV}-6.679840088}{2.374459982},
\end{aligned}
\end{equation}

\noindent We are able to model our data with two perceptron layers with two inputs $\tilde{n}_{\mathrm{I}}, \tilde{T}$ and one output $P$. This yields  

\begin{equation}
    P = \left(0.1407320023\times K_2 + 10.61859989\right)\times10^{-6} \mathrm{\frac{MeV}{fm^3}},
\end{equation}

\noindent where in perceptron layer 1
\begin{equation}
\begin{aligned}
    K_{11}&=\mathrm{tanh}\left(-1.16413+0.56553\times \tilde{n}_{\mathrm{ion}}-0.564791\times \tilde{T}\right),\\
    K_{12}&=\mathrm{tanh}\left(0.720929-0.365988\times \tilde{n}_{\mathrm{ion}}-0.398236\times \tilde{T}\right),
\end{aligned}
\end{equation}

\noindent and in perceptron layer 2

\begin{equation}
\begin{aligned}
    K_2 = 0.0768876-1.56601\times K_{11}-2.21633\times K_{12}.
\end{aligned}
\end{equation}

\noindent Note $\tanh (x)=2 \sigma(2 x)-1$. For this exercise we have used 126 samples, our learning rate is 0.01 and maximum relative error is 1.91 \% using a Quasi-Newton method to train the ANN.


\bsp	
\label{lastpage}
\end{document}